\documentstyle[epsfig,here,12pt]{article}

\input amssym.def       
\input amssym.tex

\def\double{\Bbb}

\def\cc{{\double C}}

\def\rr{{\double R}}     
\def\kk{{\double K}}

\def\aa{{\cal A}}

\def\dd{{\cal D}}
\def\ee{{\cal E}}
        
\def\hh{{\cal H}}
\def\hhh{{\double H}}   
\def\ll{{\cal L}}
\def\mm{{{\cal M}}}    

\def\jj{{\cal J}}
\def\rrr{{\cal R}} 
\def\ss{{\cal S}}

\def\t{\mathrm{Tr}}

\def\lb{\left[} 
\def\rb{\right]}
\def\lp{\left(} 
\def\rp{\right)}

\def\ov{\overline}
\def\ot{\otimes}
\def\op{\oplus}
\def\om{\ominus}

\def\bbb{\begin{eqnarray}}
\def\eee{\end{eqnarray}}
\def\pp{\pmatrix}
\def\n{\nonumber}

\begin{document}

\hsize 17truecm
\vsize 24truecm
\font\twelve=cmbx10 at 13pt
\font\eightrm=cmr8
\baselineskip 16pt

\begin{titlepage}

\centerline{\twelve CENTRE DE PHYSIQUE THEORIQUE}
\centerline{\twelve CNRS - Luminy, Case 907}
\centerline{\twelve 13288 Marseille Cedex 9}
\vskip 2truecm
\begin{center}
{\bf\Large  \sc CONSTRAINTS ON SCALAR POTENTIAL\\
\bigskip
FROM SPECTRAL ACTION PRINCIPLE}
\end{center}
\bigskip
\begin{center}
{\bf Thomas KRAJEWSKI}
\footnote{ and Universit\'e de Provence and Ecole Normale Superieure de Lyon, tkrajews@cpt.univ-mrs.fr} \\
\end{center}
\bigskip
\vskip 2truecm
\leftskip=1cm
\rightskip=1cm
\centerline{\bf Abstract} 
\medskip
Using noncommutative geometry, the standard tools of differential geometry can be extended to a broad class of spaces whose coordinates are noncommuting operators acting on a Hilbert space. In the simplest case of coordinates being matrix valued functions on space-time, the standard model of particle physics can be reconstructed out of a few basic principles. Following these ideas, we investigate the general case of models arising from matrices and give the resulting constraints  on the scalar potential and gauge couplings constants, as well as some relations between fermionic and bosonic masses.  

\bigskip
PACS-92: 11.15 Gauge field theories\\ 
\indent
MSC-91: 81T13 Yang-Mills and other gauge theories\\ 
\vskip 1truecm

\noindent April 1998\\
\vskip 0.2truecm
\noindent
CPT-97/P.3523\\
\end{titlepage}

\section{Introduction}

It is nowadays well admitted that the major dificulty in quantizing gravity lies in our current concepts of geometry. Indeed, at very small scale, a simple quantum mechanical argument combined with general relativity shows that the standard notion of point cannot hold \cite{dop}. Accordingly, the usual notions of geometry must be reformulated without any reference to points.
 
\par 

In quantum mechanics, the classical algebra of observables, that are complex valued functions on  the phase space, is replaced by a noncommutative algebra. The commutation relations of this algebra lead to Heisenberg's uncertainty relations and thus to the disappearence of points. This is the basic idea of noncommutative geometry: the standard algebra of functions on a space must be replaced by a more general noncommutative algebra, while extending the standard definitions and theorems of classical differential geometry. The most basic example of such a noncommutative geometry is certainly the quantum plane, whose coordinates satisfy $xy=qyx$ instead of $xy=yx$. On such a space, a differential calculus can be constructed and its symmetries are properly described by a quantum group $SL_{q}(2)$.  

\par

Following the analogy with quantum mechanics, one can also build a noncommutative geometry by replacing the algebra of coordinates by a noncommutative algebra represented as operators acting on a given Hilbert space. This is the setting of the theory developped by Connes, which is based on spectral triples, and allows to generalize to a broad class of spaces the notion of spin geometry \cite{book}. A spectral triple $(\aa,\hh,\dd)$ consists in an involutive algebra, playing the role of the algebra of coordinates, which is represented on the Hilbert space $\hh$. The latter is the analogue of the fermionic Hilbert space and the operator $\dd$ is the Dirac operator. To provide a suitable extension of usual geometry, the previous objects are submitted to some relations called axioms of noncommutative geometry \cite{grav}. These basic data provides us with an extension of standard Yang-Mills theory, keeping valid, for instance, the analogue of the Atiyah Singer index theorem. 

\par

Although spectral triples generalize Yang-Mills theories, including the standard model with its symmetry breaking potential,  they do not describe its coupling to gravity. The latter may be obtained by means of the spectral action principle \cite{spec}-\cite{rov}. Indeed, when the algebra of coordinates is chosen to be the algebra of functions on space-time with value in a particular matrix algebra, one can reconstruct the lagrangian of the standard model coupled to gravity. Moreover, this idea can be applied in various situations, including superstring theories \cite{ali}. Here, we will be interested in studying the general features of models arising from matrix valued functions on space-time.

\par

This paper is divided into three parts. First we study the particular case of finite spectral triples, caracterized by finite dimensional algebras of coordinates. The latter describe the internal spaces of Yang-Mills theories coupled to symmetry breaking scalars. In particular, the scalars are interpreted as connections on a discrete space, described by lattice, and the potential is obtained as a sum over all its closed loops. In a second part, we combine finite spectral triples with the usual geometry of space-time and derive the action of the underlying Yang-Mills-Higgs model coupled to gravity, with special emphasis on the resulting constraints on masses and coupling constants. 
Lastly, we illustrate the previously introduced methods on a variant of the standard model, motivated by a possible quantum group symmetry.

\section{Spectral action for finite spectral triples}

Before committing ourselves in a discussion of higher dimensional models, let us have a closer look at finite spectral triples \cite{kra}-\cite{pas}. The latter are spectral triples whose algebra of coordinates and Hilbert space are finite dimensional and provide us with the simplest examples of noncommutative geometry. 

\par

In this case, the algebra $\aa$ is a direct sum of $N$ matrix algebras $M_{n_{i}}(\kk)$ whose entries are real, complex or quaternionic. To simplify our discussion, we assume that they are complex numbers and we will extend our results to the other cases at the end of this section. Thus, the gauge group is, up to some abelian factors, the direct product of the simple unitary groups $SU(n_{i})$. Moreover, the representation of $\aa$ on the Hilbert space is supposed to be linear over complex numbers, so that the Hilbert space can be  
decomposed as
$$
\hh=\mathop{\op}\limits_{1\leq i,j\leq N}\hh_{ij},
$$
where fermions in $\hh_{ij}$ transform as the tensor product $n_{i}\ot\ov{n_{j}}$ of two fundamental representations. These gauge multiplets are chiral fermions and we represent them as the vertices of a lattice, putting at the point of coordinate $(i,j)$ a vertex of type $\om$ if fermions of $\hh_{ij}$ are left-handed and one of type $\op$ if they are right-handed. The charge conjugation $\jj$ is an antilinear map that exchanges $\hh_{ij}$ and $\hh_{ji}$. The non vanishing matrix elements of the Dirac operator $\dd$ between various subspaces give rise to links relating the corresponding vertices; whose physical counterparts are Yukawa couplings. The axioms of noncommutative geometry require that the resulting diagram be symmetric with respect to the diagonal and be made only of vertical and horizontal links relating vertices of different type.

\par

In general, a unitary element $u$ of $\aa$ acts on spinors as $u\jj\;u\jj^{-1}$. Thus, the fermionic action is gauge invariant if we replace the free Dirac operator $\dd$ by the covariant one $\dd+A+\jj A\jj^{-1}$. In the previous relation, $A$ denotes the gauge field, that is, a self adjoint operator that can be written as
$$
A=\mathop{\sum}\limits_{i}a_{0}^{i}\lb\dd ,a_{1}^{i}\rb,
$$
where $a_{o}^{i}$ and $a_{1}^{i}$ are elements of the algebra. Under a gauge transformation, $A$ transforms in the usual manner,
$$
A\rightarrow uAu^{*}+u\lb\dd,u^*\rb,
$$
so that
$$
\dd+A+\jj A\jj^{-1}\rightarrow 
u\jj u\jj^{-1}\lp\dd+A+\jj A\jj^{-1}\rp u^{*}\jj u^{*}\jj^{-1},
$$
which ensures gauge invariance of the fermionic action. For finite spectral triples, the Dirac operator can be written as
$$
\dd=\Delta+\jj\Delta\jj^{-1},
$$
where $\Delta$ is a gauge field. This splitting of $\dd$ allows us to rewrite the covariant Dirac operator as
$$
\dd+A+\jj A\jj^{-1}=\Phi+\jj\Phi\jj^{-1},
$$
with $\Phi=\Delta+A$ in the space of gauge fields. After addition of space-time dependence, $\Phi$ will be identified with the scalar field and its transformation law is simply
$$
\Phi\rightarrow u\Phi u^{*}.
$$
However, $\Phi$ is a large sparse matrix and only the non vanishing blocks that are full matrices can be identified with scalar fields. Accordingly, we must rewrite $\Phi$ in terms of smaller matrices that are the true scalars of the theory and give their transformation laws. To proceed, let us introduce the orthogonal projection $P_{ij}$ from $\hh$ onto $\hh_{ij}$. It can be shown \cite{kra} that
$$
\phi=\mathop{\sum}\limits_{ijkp}
P_{ik}^{*}\lp\Phi_{ij}^{p}\ot M_{ij,k}^{p}\ot I_{n_{k}}\rp P_{jk},
$$
where $M_{ij,k}^{p}$ are generalized mixing matrices between fermionic generations. $\Phi_{ij}^{p}\in M_{n_{i}\times n_{j}}(\cc)$ are scalar fields transforming as
$$
\Phi_{ij}^{p}=u_{i}\Phi_{ij}^{p}u_{j}^{*},
$$
with $u_{i}\in U(n_{i})$ (the group of unitary matrices of $M_{n_{i}}(\cc)$) and $u_{j}\in U(n_{j})$. The index $p$ differentiates distinct  scalars having the same transformation law. It is worth noticing that we never have $i=j$, so that the scalar fields never sit in the adjoint representation of the simple groups appearing in the gauge group. Since $\Phi$ is hermitian we have $\lp\Phi_{ij}^{p}\rp^{*}=\Phi_{ji}^{p}$. The same decomposition holds for $\jj\Phi\jj^{-1}$,
$$
\jj\Phi\jj^{-1}=\mathop{\sum}\limits_{ijkp}
P_{ki}^{*}\lp I_{n_{k}}\ot \ov{M}_{ij,k}^{p}\ot \ov{\Phi}_{ij}^{p}\rp P_{kj},
$$
where the bar stands for complex conjugation.

\par

In the commutative case, the algebra is just a finite number of copies of $\cc$ and corresponds to the algebra of functions on a finite set of points. Within this picture, the complex field $\Phi_{ij}^{p}$ appears as a connection between the points $i$ and $j$. In the noncommutative case, the algebra $\aa$ is interpreted as the algebra of endomorphisms of the space of sections of a "vector bundle" over the previous finite set, that is, we simply attach to each point a given finite dimensional vector space, whose dimension can vary from point to point. Thus the matrix $\Phi_{ij}^{p}$ still can be interpreted as a connection between fibers lying over points $i$ and $j$.  

\par

The spectral action principle states that the action can only depend on the spectrum of the covariant Dirac operator. In our case, the latter is just the matrix $\Phi+\jj\Phi\jj^{-1}$. After incoporation of the space-time geometry, only functions like
$$
S\lb\Phi\rb=\mathop{\sum}\limits_{n=0}^{\infty}
\frac{1}{\Lambda^{n}}\; a_{n}\;\t\lp\lp\Phi+\jj\Phi\jj^{-1}\rp^{n}\rp.
$$
will appear in the heat kernel development. We have introduced the scale $\Lambda$ so that $\Phi /\Lambda$ be dimensionless. Note that all odd terms vanish since $\Phi$ anticommutes with the chirality. When properly expressed in terms of the matrix valued scalar fields $\Phi_{ij}^{p}$ and after incorporation of space-time geometry, the previous power expansion will give us the scalar potential of the resulting Yang-Mills-Higgs model.

\par   

To proceed, let us introduce
$$
\Phi_{ij}^{kl}=P_{ij}^{*}\lp \Phi+\jj\Phi\jj^{-1}\rp P_{kl}.
$$
The trace of the 2n-th power of $\Phi+\jj\Phi\jj^{-1}$ can be rewritten as
$$
\t\lp\lp\phi+\jj\Phi\jj^{-1}\rp^{2n}\rp=
\mathop{\sum}\limits_{i_{1},...,i_{2n},\,  j_{1},...,j_{2n}}
\t \lp \Phi_{i_{1}j_{1}}^{i_{2}j_{2}}\Phi_{i_{2}j_{2}}^{i_{3}j_{3}}...
\Phi_{i_{2n}j_{2n}}^{i_{1}j_{1}}\rp.
$$
The sequence of points $(i_{1},j_{1})$, $(i_{2},j_{2})$,..., $(i_{n},j_{n})$, $(i_{1},j_{1})$ determines a closed loop on the diagram given by the non vanishing matrix elements of the Dirac operator. The matrix elements $\Phi_{ij}^{kl}$ vanish if $i\neq j$ and $k \neq l$ or if $i=j$ and $k=l$. In all other cases we have
\bbb
\Phi_{ij}^{kl}&=&\mathop{\sum}\limits_{p}
P_{ij}^{*}\lp \Phi_{ik}^{p}\ot M_{ik,j}^{p}\ot I_{n_{j}}\rp P_{kl}\;\;\;
(\mathrm{for}\; j=l),\n\\
\Phi_{ij}^{kl}&=&\mathop{\sum}\limits_{p}
P_{ij}^{*}\lp I_{n_{i}} \ot \ov{M}_{jl,i}^{p}\ot \ov{\Phi}_{jl}^{p}\rp P_{kl}
\;\;\;
(\mathrm{for}\; i=k).\n
\eee
Accordingly,
\bbb
&\t\lp\phi+\jj\Phi\jj^{-1}\rp^{2n}\;\;\; =\;\;\;
\mathop{\sum}\limits_{\mathrm{loops}}\;\;\;\;
\mathop{\sum}\limits_{\mathrm{multiplicities}}&\n\\
&\t\lp M_{i_{1}i_{2},j_{1}}^{p_{1}}\ov{M}_{j_{1}j_{2},i_{2}}^{q_{1}}.....
M_{i_{r}i_{1},j_{1}}^{p_{r}}\rp
\t\lp \Phi_{i_{1}i_{2}}^{p_{1}}\Phi_{i_{2}i_{3}}^{p_{2}}....
\Phi_{i_{r}i_{1}}^{p_{r}}\rp
\t\lp \ov{\Phi}_{j_{1}j_{2}}^{q_{1}}\ov{\Phi}_{j_{2}j_{3}}^{q_{2}}....
\ov{\Phi}_{j_{s}j_{1}}^{q_{s}}\rp.&\n
\eee 
The loop made of $(i_{1},j_{1})$, $(i_{2},j_{1})$, $(i_{2},j_{2})$,.....,$(i_{r},j_{1})$, $(i_{1},j_{1})$ has length 2n and consists of $r$ vertical and $s$ horizontal links. The indices $p$ and $q$ refer to vertical and horizontal multiplicities. When we sum over loops,we must be aware that we have also to take into account trivial loops, i.e. loops that are homotopic to a point within the diagram. For instance, the sequence
$$
(i,k)\rightarrow (j,k)\rightarrow (i,k)
$$
contributes to the sum as
$$
\t\lp M_{ik,j}M_{ki,j}\rp\t\lp\Phi_{ik}\Phi_{ki}\rp
$$
which is in general non-zero. Note that the sum extends over oriented loops, two loops differing just by their orientation give rise to complex conjugate terms. Moreover the symmetry operation constisting of an exchange of vertical and horizontal lines also provides complex conjugate terms, so that the complete action is real. Although the summation over loops strengthens the similarity with what happens on the lattice, in this last case the matrices are unitary so that trivial loops do not contribute to the action.

\par

Since we are essentially interested in constructing a Yang-Mills theory with a symmetry breaking scalar potential, we detail the construction of monomials up to degree four. In this case, the general inventory of closed loops is reduced to that of the six following types of subdiagrams. Moreover, their orientation is simply taken into account by means of a combinatorial factor.

\par

The first terms we encounter are mass terms that occur as closed loops of length two. The latter are diagrams like

\begin{figure}[H]
\centering
\epsfig{file={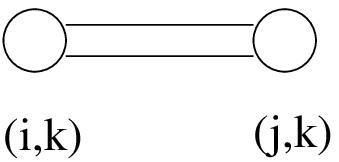},width=2.5cm}
\end{figure}

and their contribution to the mass term is
$$
2n_{k}\,
\t(M_{ij,k}M^{*}_{ij,k})\,
\t(\Phi_{ij}\Phi^{*}_{ij}).
$$
For bookeeping purposes, summation over multiplicities will be self-understood till the end of this section. Thus, the mass term just results from a summation over all edges.

\par 

The quartic self-couplings of the scalars are obtained through loops of length four. The latter are classified into the following five types that are given together with their contribution to the quartic term of the potential.

\begin{figure}[H]
\centering
\epsfig{file={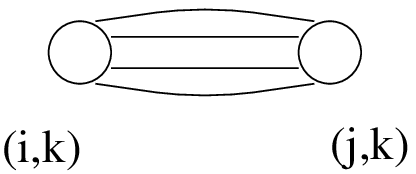},width=3cm}
\end{figure}

$$
2n_{k}\,
\t(M_{ij,k}M_{ij,k}^{*}M_{ij,k}M_{ij,k}^{*})\,
\t(\Phi_{ij}\Phi_{ij}^{*}\Phi_{ij}\Phi_{ij}^{*})
$$

\begin{figure}[H]
\centering
\epsfig{file={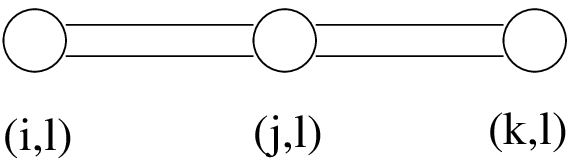},width=4.5cm}
\end{figure}

$$
4n_{k}\,
\t(M_{ij,l}M_{jk,l}M_{jk,l}^{*}M_{ij,l}^{*})\,
\t(\Phi_{ij}\Phi_{jk}\Phi_{jk}^{*}\Phi_{ij}^{*})
$$

\begin{figure}[H]
\centering
\epsfig{file={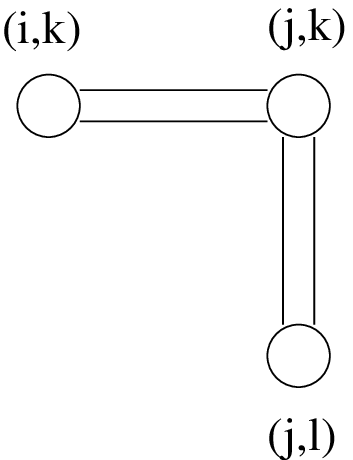},width=2.5cm}
\end{figure}

$$
4\,
\t(M_{ij,k}\ov{M}_{kl,j}\ov{M}_{kl,j}^{*}M_{ij,k}^{*})\,
\t(\Phi_{ij}\Phi_{ij}^{*})\,
\t(\Phi_{kl}\Phi_{kl}^{*})
$$

\begin{figure}[H]
\centering
\epsfig{file={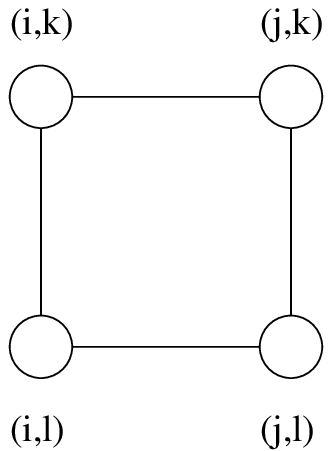},width=3cm}
\end{figure}

$$
8\,
\t(M_{ij,k}\ov{M}_{kl,j}M_{ij,l}^{*}\ov{M}_{kl,i}^{*})\,
\t(\Phi_{ij}\Phi_{ij}^{*})\,
\t(\Phi_{kl}\Phi_{kl}^{*})
$$
\begin{figure}[H]
\centering
\epsfig{file={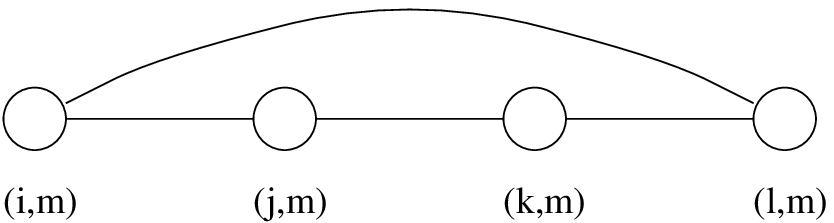},width=8cm}
\end{figure}

$$
8n_{k}\,
\t(M_{ij,m}M_{jk,m}M_{kl,m}M_{li,m})\,
\t(\Phi_{ij}\Phi_{jk}\Phi_{kl}\Phi_{li})
$$
Accordingly, the most general form of the scalar potential is
\bbb
&V(\Phi_{ij})=\mathop{\sum}\limits_{ij}\;\mu_{ij}\;
\t\lp\Phi_{ij}\Phi_{ij}^{*}\rp&\n\\
&+\mathop{\sum}\limits_{ijkl}\;\kappa_{ijkl}\;
\t\lp\Phi_{ij}\Phi_{ij}^{*}\rp\t\lp\Phi_{kl}\Phi_{kl}^{*}\rp\;
+\mathop{\sum}\limits_{ijkl}\;\lambda_{ijkl}\;
\t\lp\Phi_{ij}\Phi_{jk}\Phi_{kl}\Phi_{li}\rp.&\n
\eee
From the lattice gauge theory viewpoint, the last terms correspond to what is called "plaquette interactions" and is certainly the most difficult to work with, because it involves the trace of four different fields.

\par

In general, given scalar fields $\Phi_{ij}$, with $i\neq j$, transforming under a gauge transformation as
$$
\Phi_{ij}\rightarrow u_{i}\,\Phi_{ij}\, u_{j}^{-1},
$$
the most general, gauge invariant, potential that can be written is a sum of products of "plaquette interactions" like
$$
\t\lp\Phi_{i_{1}i_{2}}\Phi_{i_{2}i_{3}}...\Phi_{i_{n}i_{1}}\rp,
$$
where $i_{n}\neq i_{n-1}$. Thus, irrespectively of its coefficients, the scalar potential of degree at most 4 arising from the spectral action principle contains the most general gauge invariant monomials. However, we can only construct monomials that are at most products of two "plaquette interactions", so that the following gauge invariant term of degree six is forbidden in noncommutative geometry:
$$
\t\lp\Phi_{i_{1}i_{2}}\Phi_{i_{1}i_{2}}^{*}\rp
\t\lp\Phi_{j_{1}j_{2}}\Phi_{j_{1}j_{2}}^{*}\rp
\t\lp\Phi_{k_{1}k_{2}}\Phi_{k_{1}k_{2}}^{*}\rp.
$$

\par

The previous construction must also be compared with the Connes-Lott model \cite{cl} whose most general scalar potential has been derived in \cite{ali'}. Since the Connes-Lott models have the same scalar field content as those based on the spectral action principle, and since both are quartic, it follows from the previous discussion that their general form must agree. Thus, only numerical coefficients can differ. However, these two models are quite different in spirit. Indeed, the Connes-Lott model was based on generalization of the differential calculus and the Higgs potential was the analogue of the quartic Yang-Mills action. On the contrary, the spectral action principles emphasizes the analogy with general relativity, and the Higgs field is rather considered as a generalization of the Dirac operator, its quartic potential being obtained through a heat-kernel expansion. In particular, two major differences are worthwhile being noticed. First of all, in the Connes-Lott model, we always start with the fermionic mass matrix, compute the Higgs field and then its curvature and its quartic potential, whose true minimum always proves to be the mass matrix we started with. In the model we study, this never happens in general. The matrix we started with only enters the game through the matrices $M_{ij,k}^{p}$ and it is in general impossible to determine the vacuum esxplicitely in terms of the input. Secondly, multiplying all matrices $M_{ij,k}$ by a constant factor does not affect the result, contrary to the mass matrix of the Connes-Lott model. Therefore, a scale $\Lambda$ must be introduced by hand. 

\par

All the recipes we developped in this section most conveniently work in the complex case, i.e. when the algebra and its representation are complex ones. However, they may be  extended to the real case, provided minor modifications are done. Firstly, the indices $i,j,k,...$ must be thought as labelling the irreducible real representations of the simple factors of the algebra instead these factors in itself, so that it allows for the fundamental representation of $M_{n}(\cc)$ as well as its complex conjugate. Secondly, when computing the scalar fields, one encounters relations like $\Phi_{i\ov{j}}=\ov{\Phi}_{ij}$, where the representation $\ov{j}$ denotes the complex conjugate of $j$, and $\ov{\Phi_{ij}}$ the complex or quaternionic conjugate of $\Phi_{ij}$. Finally, some of the terms given by the previous loop expansion vanish identically. Indeed, the scalar product of a complex doublet and its quaternionic conjugate always vanishes, as can be seen in the case of the standard model.

\section{Constraints on higher dimensional models}

To construct a Yang-Mills model with spontaneous symmetry breaking coupled to gravity, we must incorporate the geometry of the space-time. To proceed, we simply take the product of space-time by a finite noncommutative space described by the finite spectral triple $(\aa_{F},\hh_{F},\dd_{F})$. This amouts to define a new spectral triple $(\aa,\hh,\dd)$ by
\bbb
\aa &=& C^{\infty}(\mm)\ot\aa_{F},\n\\
\dd &=& i\gamma^{\mu}(\partial_{\mu}+\omega_{\mu})\ot 1
+\gamma^{d+1}\ot \dd_{F}, \n\\
\hh &=& \ss\ot\hh_{F},\n
\eee  
where space-time is taken to be a compact and Riemannian manifold $\mm$ of even dimension $d$ with spin structure, $C^{\infty}(\mm)$ denotes the smooth complex valued functions on $\mm$ and $\ss$ is the Hilbert space of square integrable spinors on $\mm$. The operator $i\gamma^{\mu}(\partial_{\mu}+\omega_{\mu})$ is the standard Dirac operator that will be described below in more details and $\gamma^{d+1}$ stands for the d-dimensional chirality. This spectral triple is actually a graded tensor product of spectral triples, so that the axioms hold as soon as they hold for each factor. Note that the algebra of coordinates is just obtained by taking matrix valued functions instead of complex functions. Matrix valued functions correspond to the endomorphisms of a trivial vector bundle over $\mm$, whereas non trivial vector bundles give rise to more complicated algebras that are not a tensor product of complex functions by a matrix algebra. However, this case can as well be treated within the framework of noncommutative geometry, provided minor changes are made. 

\par 

As usual, fermions are coupled to general relativity through the introduction of the vielbeins $ e_{a}=e_{a}^{\mu}\partial_{\mu}$. The matrix $e_{a}^{\mu}$ describes the transition from the coordinate basis to the non-coordinate one, in which the metric tensor is euclidean. Thus $g^{\mu\nu}=e_{a}^{\mu}e_{a}^{\nu}$. The latin indices $a$, $b$, $c$,... are reserved to the non-coordinate basis and $\kappa$, $\lambda$, $\mu$, $\nu$,... are usual space-time indices.The curved space Dirac matrices are defined as $\gamma^{\mu}=e_{a}^{\mu}\gamma^{a}$, where $\gamma^{a}$ denote the euclidean Dirac matrices chosen to be hermitian. The spin connection $\omega_{\mu}=1/4\, \omega_{ab\mu}\gamma^{ab}$, with $\gamma^{ab}=1/2\lb\gamma^{a},\gamma^{b}\rb$, is nothing but the Levi-Civita connection with two indices in the non-coordinate basis. Accordingly, it must fulfill, in the torsion free case,
$$
\partial_{\mu}\, e^{\nu}_{a}+\omega_{ab\mu}\, e^{\nu}_{b}+
 \Gamma^{\nu}_{\lambda\mu}\, e^{\lambda}_{a}=0,
$$
where $\Gamma^{\kappa}_{\mu\nu}$ are the Christoffel symbols.

\par

Under a gauge transformation, a spinor $\Psi$ is transformed into $u\;\jj u\jj^{-1}\;\Psi$, where $u$ is a unitary element of $\aa$ and $\jj$ is the Tomita operator of the finite spectral triple. Its space-time analogue, the standard charge conjugation, does not play any role in this transformation law, since it commutes with the action of $\aa$ and with the Dirac operator. Covariance under these transformations leads us to introduce the gauge field $A$, which is a hermitian operator that can be written as
$$
A=\mathop{\sum}\limits_{i}a_{0}^{i}\lb\dd ,a_{1}^{i}\rb,
$$
with $a_{0}^{i},a_{1}^{i}\in\aa$. Accordingly, $\dd+A+\jj A\jj^{-1}$ transforms homogeneously and can be rewritten as
$$
\dd_{A,\Phi}=i\gamma^{\mu}\lp\partial_{\mu}+\omega_{\mu}+\tilde{A}_{\mu}\rp
+\gamma^{d+1}\,\tilde{\Phi},
$$
Where $\tilde{A}_{\mu}=A_{\mu}+\jj A_{\mu}\jj^{-1}$ is just a genuine gauge field that we choose to be antihermitian and $\tilde{\Phi}=\Phi+\jj\Phi\jj^{-1}$
is the scalar field we studied in the previous section.

\par

Following the spectral action principle, we construct an action that depends only on the eigenvalues of the covariant Dirac operator. Such an action is 
$$
S_{\Lambda}\lb e_{\alpha}^{\mu},A_{\mu},\Phi\rb=
\t\lp\mathrm{F}\lp \dd_{A,\Phi}/\Lambda\rp^{2}\rp,
$$
where F is a function that will be precised below and $\Lambda$ is a scale that will play the role of a cut-off and will be taken to be of order of the Planck scale. Note that the basic variables are the Yang-Mills field, the scalar field and the vielbein. Since we are in even dimension, the spectrum of $\dd$ is even and nothing is lost when we take a function of its square instead of a function of $\dd$. The previous action must be understood as an asymptotic expression when the cut-off $\Lambda$ goes to infinity.

\par

To developp the previous expression in inverse powers of $\Lambda$ we use the heat-kernel expansion. To proceed, we have to split the square of the Dirac operator into a generalized Laplacian and an endomorphism of the underlying vector bundle. Recall that if $\ee$ is the space of section of a vector bundle and $\nabla:\;\ee\,\rightarrow\,\ee\ot C^{\infty}(\mm)$ is a connection given in a chart by
$$
\nabla(\Psi)=\nabla_{\mu}(\Psi)\ot dx^{\mu},
$$
the generalized Laplacian associated to $\nabla$ is given by
$$
\Delta(\Psi)=g^{\mu\nu}\nabla_{\mu}\nabla_{\nu}(\Psi)
-g^{\lambda\mu}\Gamma_{\lambda\mu}^{\nu}\nabla_{\nu}(\Psi).
$$
Since the square of the Dirac operator is a second order differential operator whose leading symbol is given by the opposite of the metric tensor, it is known that there is a unique connection $\nabla$ such that $\dd^{2}=-\Delta+E$, where $\Delta$ is the generalized Laplacian associated to $\nabla$ and $E$ an endomorphism of the bundle \cite{gil}. This decomposition is obtain as follows.
\bbb
\dd^{2}&=&
\lp i\gamma^{\mu}\nabla_{\mu}+\gamma^{d+1}\tilde{\Phi}\rp
\lp i\gamma^{\nu}\nabla_{\nu}+\gamma^{d+1}\tilde{\Phi}\rp\n\\
&=&-\gamma^{\mu}\nabla_{\mu}\gamma^{\nu}\nabla_{\nu}
+i\gamma^{\mu}\gamma^{d+1}D_{\mu}\tilde{\Phi}+\tilde{\Phi}^{2}\n\\
&=&-\gamma^{\mu}\gamma^{\nu}\nabla_{\mu}\nabla_{\nu}
-\gamma^{\mu}\lb\partial_{\mu}+\omega_{\mu},\gamma^{\nu}\rb
\nabla_{\nu}\n\\
& &+i\gamma^{\mu}\gamma^{d+1}D_{\mu}\tilde{\Phi}+\tilde{\Phi}^{2}\n\\
&=&-g^{\mu\nu}\nabla_{\mu}\nabla_{\nu}
+g^{\mu\nu}\Gamma_{\mu\nu}^{\lambda}
\nabla_{\lambda}\n\\
& &-\frac{1}{2}\gamma^{\mu\nu}\Omega_{\mu\nu}
+i\gamma^{\mu}\gamma^{d+1}D_{\mu}\tilde{ \Phi}+\tilde{\Phi}^{2}\n
\eee
The first line gives us the opposite of the generalised Laplacian $\Delta$ and the second one the endomorphism $E$. We have introduced 
\bbb
\nabla_{\mu}&=&\partial_{\mu}+\omega_{\mu}+\tilde{A}_{\mu}\n\\
D_{\mu}\tilde{\Phi}&=&\partial_{\mu}\Phi+\lb \tilde{A}_{\mu},\tilde{\Phi}\rb,\n\\
\gamma^{\mu\nu}&=&\frac{1}{2}\lb\gamma^{\mu},\gamma^{\nu}\rb,\n\\
\Omega_{\mu\nu}&=&\lb\nabla_{\mu},\nabla_{\nu}\rb.\n\\
&=&\frac{1}{4}R_{ab\mu\nu}\gamma^{ab}+\tilde{F}_{\mu\nu},\n\\
\tilde{F}_{\mu\nu}&=&\partial_{\mu}\tilde{A}_{\nu}-\partial_{\nu}\tilde{A}_{\mu}
+\lb \tilde{A}_{\mu},\tilde{A}_{\nu}\rb,\n
\eee
and where $R_{ab\mu\nu}$ denotes the Riemann tensor. The two-form $\tilde{F}_{\mu\nu}$ is the curvature of the gauge field $\tilde{A}_{\mu}$. Note that $\Omega_{\mu\nu}$ is nothing but the curvature of the connection $\nabla_{\mu}$ involved in the definition of the generalized Laplacian. It is also worth noticing that the vielbein has been completely eliminated, so that the gravitational sector of the action only depends on the metric tensor. Further contraction of $R_{ab\mu\nu}$ yields
$$
\gamma^{\mu\nu}R_{ab\mu\nu}\gamma^{ab}=2\rrr,
$$
so that, as a final result, $E$ is given by
$$
E=-\frac{1}{4}\rrr-\frac{1}{2}\gamma^{\mu\nu}\tilde{F_{\mu\nu}}
+i\gamma^{\mu}\gamma^{d+1}
D_{\mu}\tilde{\Phi}+\tilde{\Phi}^{2}.
$$

\par

Owing to the spectral action principle, the action is written as $S=\t\,F(\dd^{2}/ \Lambda^{2})$. The scale $\Lambda$ is assumed to be of order of the Planck mass and we will be interested in the asymptotic behavior of $S$ when the scale goes to infinity. The latter is given, using the heat-kernel expansion, by
$$
S\simeq\mathop{\sum}\limits_{n=0}^{d}\;F_{n}\Lambda^{d-n}\;
\int_{\mm}\; dv\; a_{n},
$$
up to terms that vanish in the limit $\Lambda\rightarrow\infty$. The coefficients $F_{n}$ only depend on the function $F$ and are given by
$$
F_{0}=\int_{0}^{\infty}tF(t)dt,\;\; 
F_{2}=\int_{0}^{\infty}F(t)dt,\;\; 
F_{2n}=(-1)^{n}F^{(n-2)}(0)\;\;
\mathrm{for}\; n\geq 2.
$$
Accordingly, the lagrangian derived from the spectral action is
$$
\ll_{\Lambda}\lp g_{\mu\nu}, A_{\mu},\Phi\rp=\mathop{\sum}\limits_{n=0}^{d}\;F_{n}\Lambda^{d-n}\;a_{n}.
$$ 
The coefficients $a_{n}$ vanish for odd values of $n$. In the even case, they only depend on the metric tensor $g_{\mu\nu}$, on the connection $\nabla_{\mu}$ and on the endomorphism $E$. The first three non vanishing ones are given, discarding total derivatives, by

\bbb
a_{0}&=&\frac{\t(1)}{(2\pi)^{\frac{d}{2}}}\n\\
a_{2}&=&-\frac{1}{12}\;\frac{\t(1)}{(2\pi)^{\frac{d}{2}}}\;\rrr
-\frac{1}{(2\pi)^{\frac{d}{2}}}\;\t(\tilde{\Phi}^{2})\n\\
a_{4}&=&\frac{1}{1440}\;\frac{\t(1)}{(2\pi)^{\frac{d}{2}}}\;
\lp 5\rrr^{2}-8 R_{\mu\nu}R^{\mu\nu}
-7 R_{\kappa\lambda\mu\nu}R^{\kappa\lambda\mu\nu}\rp\n\\
&&+\frac{1}{2}\;\frac{1}{(2\pi)^{\frac{d}{2}}}\;
\lp \t(D_{\mu}\tilde{\Phi}D^{\mu}\tilde{\Phi})
+\t(\tilde{\Phi}^{4})\rp\n\\
&&-\frac{1}{6}\;\frac{1}{(2\pi)^{\frac{d}{2}}}\;
\t(\tilde{F}_{\mu\nu}\tilde{F}^{\mu\nu})
+\frac{1}{12}\;\frac{1}{(2\pi)^{\frac{d}{2}}}\;
\rrr\t(\tilde{\Phi}^{2}),\n
\eee

$\t(1)$ denoting the dimension of the finite dimensional Hilbert space $\hh_{F}$. When expressed using the matrix of multiplicities $m$ \cite{kra}, we have
$$
\t(1)=\mathop{\sum}\limits_{ij}\; m_{ij}n_{i}n_{j}.
$$
We recall that the matrix of  multiplicities classifies the bimodule structure of the finite spectral triple and it contains all information pertaining to the fermionic representation. The higher degree coefficients $a_{n}$ contain higher order derivatives and certainly lead to unphysical theories. Consequently, when the dimension is bigger that four, we will assume that the function $F$ is chosen such that these terms do not appear. In other words, we will require that all the derivatives of $F$ vanish in $0$, or, for simplicity, that $F$ is constant on a neighbourhood of the origin \cite{asymp}. In dimension four, these troublesome terms do not appear and the function $F$ can, a priori, be arbitrary. From now on, we will commit ourselves in a detailed study of the lagrangian, with special emphasis on the non trivial relations arising between the various coupling constants.

\par

Let us first investigate the constraints on the gauge coupling. The pure Yang-Mills action arises from the term $a_{4}$ and it is equal to
$$
-\frac{F_{4}\;\Lambda^{d-4}}{6(2\pi)^{\frac{d}{2}}}\t(\tilde{F}_{\mu\nu}\tilde{F}^{\mu\nu}).
$$
The antihermitian element $A_{\mu}$ of the algebra can be written as a direct sum of non-abelian gauge fields $g_{i}\, A_{\mu}^{i}\in{\frak s}{\frak u}(n_{i})$ and of the corresponding abelian ones $ B_{\mu}^{i}\in{\frak u}(1)=i\rr$, where the positive real numbers $g_{i}$ stand for the non-abelian coupling constants. The coupling constants of the abelian part will be introduced later, after application of a unimodularity condition \cite{cl}. Accordingly, we have
$$
\t(\tilde{F}_{\mu\nu}\tilde{F}^{\mu\nu})=\mathop{\sum}\limits_{ij}
g_{i}^{2}2m_{ij}n_{j}\;\t(F_{\mu\nu}^{i}F^{i\mu\nu})
+q_{ij}\; G_{\mu\nu}^{i}G^{j\mu\nu},
$$
where $F_{\mu\nu}^{i}$ and $G_{\mu\nu}^{i}$ are the field strength tensors associated to $A_{\mu}^{i}$ and $B_{\mu}^{i}$. The symmetric matrix $q_{ij}$ defines the abelian Yang-Mills action and is given by
$$
q_{ij}=2\lp\mathop{\sum}\limits_{k}m_{ik}n_{i}n_{k}\rp\delta_{ij}-
2m_{ij}n_{i}n_{j}.
$$
Usually, the non-abelian pure Yang-Mills lagrangian for the fields $A_{\mu}^{i}$ is given by
$$
-\frac{1}{2}\mathop{\sum}\limits_{i}\,\t(F_{\mu\nu}^{i}F^{i\mu\nu}).
$$
Identification of the non-abelian Yang-Mills action arising from the spectral action principle with the previous one forces us to properly normalize the gauge fields. This yields the following expressions for the coupling constants
$$
g_{i}=(2\pi)^{\frac{d}{4}}\sqrt{\frac{3/2}{F_{4}\Lambda^{d-4}\mathop{\sum}\limits_{j}m_{ij}n_{j}}}.
$$
For the standard model, it is easily shown that this relation yields the standard expression for the nonabelian gauge coupling which are both equal at the scale $\Lambda$. However, in general this is not the case and there is no obvious relations between the gauge couplings. To tackle the abelian case, let us express the gauge fields $B_{\mu}^{i}$ as
$$
B_{\mu}^{i}=\mathop{\sum}\limits_{j=1}^{N'}P_{ij}C_{\mu}^{j}.
$$
This amounts to parametrize the space of abelian gauge fields $B_{\mu}^{i}$ by $N'\leq N$ new fields $C_{\mu}^{i}$. The real coefficients $P_{ij}$, encoding both the coupling constants and the fermionic charges, are chosen such that the abelian part of the Yang-Mills lagrangian identifies with the usual one, given by
$$
-\mathop{\sum}\limits_{i=1}^{N'}\,\frac{1}{4}\,H_{\mu\nu}^{i}H^{i\mu\nu},
$$  
where $H_{\mu\nu}^{i}$ is the curvature of $C_{\mu}^{i}$. In other words, the rectangular matrix $P_{ij}$ reduces the matrix of quadratic form $q_{ij}$ to a scalar matrix of size $N'$. It is worthwhile to note that in general, one can have $N'\leq N$, for at least two reasons. First of all, although the matrix $q_{ij}$ is negative, it can have zero modes. The latter do not appear in the action and are hence unphysical. Secondly, it often happens that some of the   abelian fields are to be ruled out by hand for physical reasons, such as anomaly cancellation \cite{jmgb}. This operation, called {\it unimodularity condition} can be achieved in this manner. In general, the matrix $P_{ij}$ has $NN'$ entries and since it reduces $q_{ij}$ to a scalar $N'\times N'$ matrix, it fulfills $1/2\;N'(N'+1)$ conditions. Moreover, these conditions are also satified if we multiply $P$ on the left by an orthogonal $N'\times N'$ matrix. Since this transformation just corresponds to a rotation in the $N'$ dimensional space of the gauge fields $C_{\mu}^{i}$, it has no physical relevance. Hence, $1/2\;N'(N'-1)$ degrees of freedom must be substracted and we end up with
$$
NN'-\frac{N'(N'+1)}{2}-\frac{N'(N'-1)}{2}=N'(N-N')
$$
arbitrary parameters in the matrix $P_{ij}$. For the standard model, we have $N=2$, because quaternions do not contribute to the abelian part of the gauge fields and $N'=1$. Consequently, there is only one degree of freedom of choosing the way the unimodularity condition is achieved. Usually, one imposes that $\t(A_{\mu})=0$, so that it fixes unambiguously all electric charges of the fermions, provided the electron charge is chosen to be minus one. 

\par

The gravitational sector of the theory consists of three different pieces. First of all, the term $\Lambda^{d}\,F_{0}\,a_{0}$ yields a constant term in the lagangian that defines the cosmological constant
$$
\Lambda_{cosm}=\frac{1}{(2\pi)^{\frac{d}{2}}}\,\t(1)\,\Lambda^{d}\,F_{0}.
$$
The second term is the usual Einstein-Hilbert lagrangian that comes from $a_{2}F_{2}\Lambda^{d-2}$. Identifying the part proportional to the scalar curvature, chosen to be positive for the spheres embedded in euclidean spaces, with
$$
-\frac{1}{16\pi G}\,\rrr
$$
yields
$$
G=\frac{3}{2}\frac{(2\pi)^{\frac{d}{2}-1}}{\t(1)}\,\Lambda^{2-d}F_{2}.
$$
It seems worthwhile to notice that the second term arising from $a_{2}$ is the mass term of the scalar, so that the latter appears on the same footing as gravitational field. At last, the coefficient $a_{4}$ provides us with a lagrangian for $\rrr^{2}$-gravity. Indeed, apart from a coupling between scalars and gravitation in $\rrr\t(\tilde{\Phi}^{2})$ and surface terms like $\Box\rrr$, the gravitational part is
$$
-\frac{\t(1)}{(2\pi)^{\frac{d}{2}}}\,F_{4}\,\Lambda^{d-4}
\lp \frac{1}{36}\rrr^{2}+\frac{1}{180}R_{\mu\nu}R^{\mu\nu}
+\frac{7}{1440}R_{\kappa\lambda\mu\nu}R^{\kappa\lambda\mu\nu}\rp.
$$
We will come back to this point when we will discuss the four dimensional case.

\par

To study the spontaneous symmetry breaking sector, we first have to properly normalize the scalar fields. The corresponding term in the action is given by
$$
\frac{1}{2}\,\frac{1}{(2\pi)^{\frac{d}{2}}}\,F_{4}\Lambda^{d-4}
\t\lp D_{\mu}\tilde{\Phi}D^{\mu}\tilde{\Phi}\rp.
$$
The kinetic part is
$$
\frac{1}{(2\pi)^{\frac{d}{2}}}\,F_{4}\Lambda^{d-4}
\t\lp\partial_{\mu}\Phi\partial^{\mu}\Phi\rp.
$$
where we have
$$
\t(\partial_{\mu}\Phi\partial^{\mu}\Phi)=
\mathop{\sum}\limits_{ijkpq}
\t\lp\partial_{\mu}\Phi_{ij}^{p}\partial^{\mu}\Phi_{ij}^{q*}\rp\; n_{k}\t\lp M_{ij,k}^{p}M_{ij,k}^{q*}\rp.
$$
This term must be identified with the usual kinetic term of the scalar fields, given by
$$
\mathop{\sum}\limits_{\mathrm{scalars}}
\frac{1}{2}\t\lp\partial_{\mu}\Phi\partial^{\mu}\Phi^{*}\rp=
\mathop{\sum}\limits_{ijp}
\frac{1}{4}\t\lp\partial_{\mu}\Phi_{ij}^{p}\partial^{\mu}\Phi_{ij}^{p*}\rp,
$$
where on the LHS the sum runs over all independent scalar fields (we recall that $\Phi_{ij}^{*}=\Phi_{ji}$, so that $\Phi_{ij}$ and $\Phi_{ji}$ are not independent fields). By construction, the matrices $M_{ij,k}^{p}$ are such that the vectors
$$
E_{p}=\lp M_{ij,1}^{p},...,M_{ij,k}^{p},...,M_{ij,N}^{p}\rp
$$
form a basis, for fixed $i$ and $j$, of a given vector space \cite{kra}. Thus, they can always be choosen to be orthogonal, by means of the Gram-Schmidt procedure. We will assume that
$$
\mathop{\sum}\limits_{k} n_{k}\t\lp M_{ij,k}^{p}M_{ij,k}^{q*}\rp = X\,\delta_{p,q},
$$
where $X$ is a positive real constant. Identification with the standard kinetic term for the scalars yields
$$
X=\frac{(2\pi)^{\frac{d}{2}}}{4F_{4}\Lambda^{d-4}}.
$$

\par

Apart from the Einstein-Hilbert lagrangian, the term containing $a_{2}$ also yields the mass term for the scalars. Indeed, the latter appears as
$$
-\frac{1}{(2\pi)^{\frac{d}{2}}}\,F_{2}\,\Lambda^{d-2}\,\t(\tilde{\Phi}^{2}).
$$
Following the diagrammatic approach developed in the previous section, $\t(\Phi)$ can be easily evaluated as a sum over all the links of the diagram. However, due to the normalization of the kinetic term for the scalars, it proves to be simply given by
$$
\t(\tilde{\Phi}^{2})=2X\;\mathop{\sum}\limits_{ijp}
\t\lp\Phi_{ij}^{p}\Phi_{ij}^{p*}\rp
$$
where $X$ is given by
$$
X=\frac{(2\pi)^{\frac{d}{2}}}{4F_{4}\,\Lambda^{d-4}}
$$
Therefore, the mass term for the scalar is
$$
-\frac{1}{2}\mu^{2}
\;\mathop{\sum}\limits_{\mathrm{scalars}}\t\lp\Phi\Phi^{*}\rp=
-\frac{1}{4}\mu^{2}
\;\mathop{\sum}\limits_{ijp}\t\lp\Phi_{ij}^{p}\Phi_{ij}^{p*}\rp.
$$
Accordingly, all the scalar fields have the same imaginary mass and it is given by
$$
-\mu^{2}=\frac{2F_{2}}{F_{4}}\Lambda.
$$
Moreover, since all the mass terms appear with the wrong sign, the corresponding scalars break the symmetry. The coupling between scalars and the gravitational field computed from $a_{4}$ is 
$$
\frac{1}{12}\;\frac{\Lambda^{d-4}F_{4}}{(2\pi)^{\frac{d}{2}}}\;
\rrr\;\t(\tilde{\Phi}^{2}).
$$
Thanks to the normalization of $\t(\tilde{\Phi}^{2})$, it simply reduces to
$$
\frac{1}{12}\;\rrr\;\mathop{\sum}\limits_{ijp}
\t\lp\Phi_{ij}^{p}\Phi_{ij}^{p*}\rp=
\frac{1}{6}\;\rrr\;\mathop{\sum}\limits_{\mathrm{scalars}}
\t\lp\Phi\Phi^{*}\rp.
$$
The quartic self-coupling terms of the scalars are obtained through $a_{4}$, these couplings can be written as
$$
\mathop{\sum}\limits_{ijklpqrs}
\kappa_{ijkl}^{pqrs}\;
\t\lp\Phi_{ij}^{p}\Phi_{ji}^{q}\rp\t\lp\Phi_{kl}^{r}\Phi_{lk}^{s}\rp
+\lambda_{ijkl}^{pqrs}\;
\t\lp\Phi_{ij}^{p}\Phi_{jk}^{q}\Phi_{kl}^{r}\Phi_{li}^{s}\rp.
$$
In the previous expression, $\kappa_{ijkl}^{pqrs}$ and $\lambda_{ijkl}^{pqrs}$
are $1/2\Lambda^{d-4}F_{4}(2\pi)^{-\frac{d}{2}}$ times the corresponding couplings computed within the diagrammatic approach.

\par

Owing to the previous discussion, it is obvious that spontaneous symmetry breaking always occurs. If we denote by $V$ the vacuum expectation value of the field $\Phi$, the fermionic mass matrix is simply given by $\tilde{V}=V+\jj V\jj^{-1}$. The mass term for the gauge bosons comes from the covariant derivative of the scalars. Replacing $\tilde{\Phi}$ by its vacuum expectation value $\tilde{V}$, it reads
$$
\frac{1}{2}\,\frac{\Lambda^{d-4}F_{4}}{(2\pi)^{\frac{d}{2}}}\,
\t\lb \tilde{A}_{\mu},\tilde{V}\rb\lb\tilde{A}^{\mu},\tilde{V}\rb.
$$
To find the spectrum of gauge fields, one has to diagonalize this quadratic form in a basis that preserves the form of the kinetic term. Although in general an explicit form of the corresponding eigenvalues cannot be given, we can find an upper bound for the masses of the gauge field. To this aim, let us denote by $M_{n}$ the real eigenvalues of the matrix $\tilde{V}$. Then, the eigenvalues of the operator $X\mapsto\lb\tilde{V},X\rb$ acting on endomorphisms of the finite dimensional Hilbert space $\hh_{F}$ are $M_{i}-M_{j}$. Thus, if we denote by $M_{f}$ the highest module of the eigenvalues of $\tilde{V}$, $M_{f}$ is nothing but the mass of the heaviest fermion and we have
\bbb
\t\lb \tilde{A}_{\mu},\tilde{V}\rb^{*}\lb\tilde{A}^{\mu},\tilde{V}\rb&=&
\t\tilde{A}_{\mu}^{*}\lb\tilde{V},\lb\tilde{V},\tilde{A}_{\mu}\rb\rb\n\\
&\leq&4M_{f}^{2}\,\t\lp \tilde{A}_{\mu}^{*}\tilde{A}^{\mu}\rp.\n
\eee
Introducing the fields $A_{\mu}^{i}$ and their abelian counterparts $C_{\mu}^{i}$, both properly normalized, we get
$$
\t\lp\tilde{A}_{\mu}^{*}\tilde{A^{\mu}}\rp=
\frac{3(2\pi)^{\frac{d}{2}}}{F_{4}\Lambda^{d-4}}
\lp\mathop{\sum}\limits_{i=1}^{N}\t\lp A_{\mu}^{i*}A^{i\mu}\rp
+\mathop{\sum}\limits_{i=1}^{N'}\frac{1}{2}C_{\mu}^{i*}C^{i\mu}\rp.
$$
Accordingly,
$$
\frac{1}{2}\,\frac{\Lambda^{d-4}F_{4}}{(2\pi)^{\frac{d}{2}}}\,
\t\lb \tilde{A}_{\mu},\tilde{V}\rb^{*}\lb\tilde{A}^{\mu},\tilde{V}\rb
\leq
6M_{f}^{2}\lp\mathop{\sum}\limits_{i=1}^{N}\t\lp A_{\mu}^{i*}A^{i\mu}\rp
+\mathop{\sum}\limits_{i=1}^{N'}\frac{1}{2}C_{\mu}^{i*}C^{i\mu}\rp
$$
When expressed using the components of the gauge field $\tilde{A}_{\mu}$ in the standard Lie algebra basis, the RHS is a quadratic form which is twice the usual mass term for the gauge field. On the other hand, the quadratic form appearing on the LHS is $6M_{f}^{2}$ the euclidean one, due to the normalization of the kinetic term. The spectrum of gauge fields is obtained by diagonalizing the LHS while preserving the RHS. Therefore, the previous inequality shows that the highest eigenvalue of the LHS is smaller than $6M_{f}^{2}$. Comparison with the usual mass term for the gauge bosons yields the inequality
$$
0\leq M_{b}^{2}\leq 3M_{f}^{2},
$$
where $M_{b}$ stands for the mass of the heaviest gauge boson of the theory. In all practical calculations, the vacuum expectation value $\tilde{V}$ depends explicitely on the scale $\Lambda$. Therefore, so do all masses appearing in the previous inequality which is valid at this scale only. To get an inequality valid at the elctroweak scale, one must the use the renormalization flow.

\par

In general, the vacuum expectation value $\tilde{V}$ is made out of the matrices $V_{ij}^{p}$ that are the vacuum expectation values of the fields $\Phi_{ij}^{p}$. The massless gauge fields correspond to the little group of $\tilde{V}$ under the action 
$$
\tilde{V}\mapsto u\jj u\jj^{-1}\;\tilde{V}\; u^{*}\jj u^{*}\jj^{-1}
$$
of the gauge group. Equivalently, this action may be given by $ V_{ij}^{p}\mapsto u_{i}V_{ij}^{p}u_{j}^{-1}$, for $u_{i}\in U(n_{i})$ and is better visualized within a diagrammatic approach. Let us take $N$ vertices corresponding to the $N$ simple factors of the algebra. Then we draw edges between the vertices $i$ and $j$ as many times as we have scalars fields $\Phi_{ij}^{p}$ carrying indices $i$ and $j$. To each of these edges corresponds a constraint $u_{i}V_{ij}^{p}u_{j}^{-1}=V_{ij}^{p}$ on the little group and one can say, roughly speaking, that the more the vertice $i$ is related to other vertices, the more the symmetry corresponding to its gauge group $U(n_{i})$ is broken. In the extremal case of an isolated vertex, corresponding to gauge fields that do not couple to the scalars, the symmetry is entirely unbroken and these gauge fields remain massless. 

\par

In general, as soon there is a non vanishing vacuum expectation value $V_{ij}^{p}$ involving the vertex $i$, the corresponding non-abelian gauge symmetry is broken. In the abelian case, the situation is rather different. Indeed, if we write the abelian part of $U(n_{i})$ as $e^{i\theta_{i}}$, the constraints from the little group are just
$$
e^{i(\theta_{i}-\theta_{j})}\;V_{ij}^{p}=V_{ij}^{p}.
$$
Equivalently, we must have $\theta_{i}=\theta_{j}$ as soon as the vertices $i$
and $j$ are related. Thus, there is a unique massless abelian field for each connected component of the diagram. Note that we have not yet applied the unimodularity condition, that may rule out these massless fields. When this condition just requires that the trace of $A_{\mu}$ vanishes, it is easily seen that this eliminates an abelian massles gauge field, since a field proportional to the identity is a zero mode of the mass matrix. As an example, let us build the diagram corresponding to the standard model.

\begin{figure}[h]
\centering
\epsfig{file={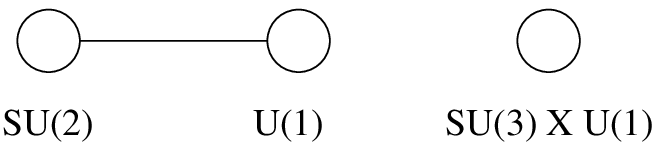},width=7cm}
\end{figure}

The only link correspond to the complex Higgs doublet. The isolated vertex is the color sector $SU(3)$ together with a extra $U(1)$ that are both massless fields. The remaining connected component stands for the electroweak sector and, according to the previous discussion, it also contains a massless $U(1)$ which is simply the little group of the vacuum expectation value of the Higgs doublet. Then, we eliminate by means of the unimodularity condition a suitable linear combination of these two massless abelian gauge fields.

\par

Finally, in the special case of a four dimensional space, the gravitational lagrangian can be simplified by introducing the Gauss-Bonnet invariant  
$$
\chi=
R_{\kappa\lambda\mu\nu}R^{\kappa\lambda\mu\nu}-4R_{\mu\nu}R^{\mu\nu}+\rrr^{2},
$$
which is a total derivative in dimension four. Indeed, given the Weyl tensor
$$
C_{\kappa\lambda\mu\nu}=R_{\kappa\lambda\mu\nu}
-\frac{1}{2}\lp g_{\mu\rho}R_{\nu\sigma}-g_{\mu\sigma}R_{\nu\rho}
+g_{\nu\sigma}R_{\mu\rho}-g_{\nu\rho}R_{\mu\sigma}\rp
+\frac{1}{6}\lp g_{\mu\rho}g_{\nu\sigma}-g_{\mu\sigma}g_{\nu\rho}\rp\rrr,
$$

whose square is
$$
C_{\kappa\lambda\mu\nu}C^{\kappa\lambda\mu\nu}=
R_{\kappa\lambda\mu\nu}R^{\kappa\lambda\mu\nu}-2R_{\mu\nu}R^{\mu\nu}
+\frac{1}{3}\rrr^{2},
$$
one has
$$
5\rrr^{2}-8R_{\mu\nu}R^{\mu\nu}-7R_{\kappa\lambda\mu\nu}R^{\kappa\lambda\mu\nu}=
-18C_{\kappa\lambda\mu\nu}C^{\kappa\lambda\mu\nu}+11\chi.
$$
Accordingly, the gravitational part of the lagrangian can be written, up to total derivatives, as
$$
\ll_{gr}=-\frac{1}{16\pi G}\rrr+
\frac{3}{160\pi\mu^{2}G^{2}}C_{\kappa\lambda\mu\nu}C^{\kappa\lambda\mu\nu}.
$$
This relates in an unexpected way Newton's coupling constant and the mass of the scalars as a coupling constant in higher derivative gravity.

\section{An example with an extension of the standard model}

Let us finally apply the previous construction to a simple example. It is conjectured that the standard model, when formulated in the framework of noncommutative geometry, may have, even at the classical level, a quantum group symmetry. This idea leads to the introduction of a finite dimensional quotient of the universal quantized envelopping algebra $U_{q}(sl_{2})$ with $q^{3}=1$. Unfortunately, within this approach it is the algebra $\cc\op M_{2}(\cc)\op M_{3}(\cc)$ which appears instead of $\cc\op\hhh\op M_{3}(\cc)$. This model provides us with an extra $U(1)$ gauge field, whose mass satisfies the inequality we derived in the previous section. Accordingly, it must be lighter than $\sqrt{3}M_{top}$, an we have to supress it by means of an another unimodularity condition \cite{sel}.

\par

To proceed, let us first recall the spectral triple of the standard model. The algebra $\aa$ is 
$$
\aa=\hhh\op\cc\op M_{3}(\cc).
$$
The Hilbert space $\hh$ is a direct sum of the particle and the antiparticle Hilbert spaces. The  particle Hilbert spaces $\hh_{L}^{P}$ and $\hh_{R}^{P}$ are spanned, for $N_{f}=3$ families of fermions, by  
$$
\pp{u\cr d}_{L},\,
\pp{c\cr s}_{L},\,
\pp{t\cr b}_{L},\,
\pp{\nu_{e}\cr e}_{L},\,
\pp{\nu_{\mu}\cr \mu}_{L},\,
\pp{\nu_{\tau}\cr \tau}_{L},
$$
and
$$
(u)_{R},\, (d)_{R},\, (c)_{R},\, (s)_{R},\, (t)_{R},
\, (b)_{R},\, (e)_{R},\, (\mu)_{R},\,(\tau)_{R},  
$$
where we have omitted the color index for quarks. The corresponding antiparticles form a basis of the antiparticle spaces $\hh_{L}^{A}$ and $\hh_{R}^{A}$. Within these bases, the representation is given by

\bbb
\pi_{L}^{P}(a)&=&\mathrm{diag}\lp a\ot I_{3N_{f}},\,a\ot I_{N_{f}}\rp\nonumber,\\
\pi_{R}^{P}(b)&=&\mathrm{diag}\lp b\,I_{3N_{f}},\,\ov{b}\,I_{3N_{f}},\,\ov{b}\,I_{N_{f}}\rp\nonumber,\\
\pi_{L}^{A}(b,c)&=&\mathrm{diag}\lp I_{2N_{f}
}\ot c,\,b\,I_{2N_{f}}\rp\nonumber,\\
\pi_{R}^{A}(b,c)&=&\mathrm{diag}\lp
I_{2N_{f}}\ot c,\,b\,I_{N_{f}}\rp\nonumber,
\eee

where $(a,b,c)\in\hhh\op\cc\op M_{3}(\cc)$. The mass matrix $M$ is 
$$
M=\pp{\pp{M_{u}\ot I_{3}&0\cr 0&M_{d}\ot I_{3}}&0\cr
0&\pp{0\cr M_{e}}},
$$
with 
\bbb
M_{u}&=&\mathrm{diag}\lp m_{u},m_{c},m_{t}\rp,\n\\
M_{d}&=&V_{CKM}\,\mathrm{diag}\lp m_{d},m_{s},m_{b}\rp,\n\\
M_{e}&=&\mathrm{diag}\lp m_{e},m_{\mu},m_{\tau}\rp,\n
\eee
where $m_{p}$ stands for the mass of particle $p$ and $V_{CKM}$ is the Cabibbo-Kobayashi-Maskawa mixing matrix.

\par

Accordingly, the matrix of multiplicities is, in the basis $(\cc,\hhh,\ov{\cc},M_{3}(\cc))$,
$$
\mu=N_{f}\,\pp{0&0&1&1\cr 0&0&-1&-1\cr 1&-1&0&1\cr 1&-1&1&0\cr},
$$
and the corresponding diagram is given by the following figure.

\par

\begin{figure}[h]
\centering
\epsfig{file={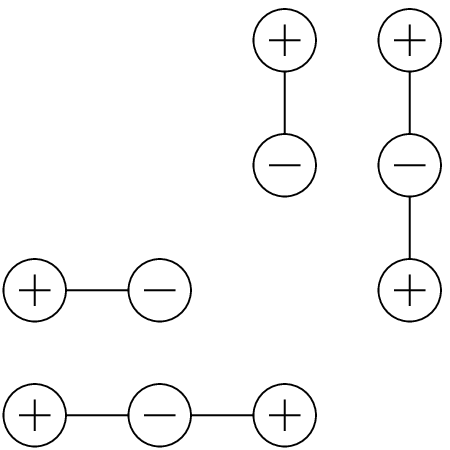},width=2.5cm}
\end{figure}

\par

Recall that to each vertex of the diagram corresponds a Hilbert space of particles or antiparticles transforming in a product of two fundamental representations of the gauge group. Particles are located above the first diagonal, and since everything is symmetric with repect to charge conjugation, we restrict our discussion to particles, multiplying the resulting potential by two.

\par

The model under consideration in this section is obtained from the standard model by replacing the algebra of quaternions by two by two complex matrices, leaving all other items of the standard model unchanged. Since this algebra is considered as a real algebra, we must distinguish the fundamental representation of a matrix algebra from its complex conjugate. Fortunately, his only happens here for the abelian factor $\cc$, and therefore we have four rows and columns in the diagram instead of the expected three. The four rows corresponds resapectively to $\ov{\cc}$, $M_{2}(\cc)$, $\cc$ and $M_{3}(\cc)$.
 
\par

The first vertical link between $M_{2}(\cc)$ and $\ov{\cc}$, the second one and the link between $M_{2}(\cc)$ and $\cc$ correspond respectively to the following matrix elements of the Dirac operator:
$$
\pp{0\cr M_{e}},\>\>\>\>\pp{0\cr M_{d}\ot I_{3}},\>\>\>\>
\pp{M_{u}\ot I_{3}\cr 0}.
$$
The first two matrix elements give rise to a doublet of complex scalar field that we denote by $\Phi_{d}$ and the last one to another doublet called $\Phi_{u}$. In the case of the standard model, these doublets are not independant, $\Phi_{u}$ being the quaternionic conjugate of $\Phi_{d}$.

\par

According to the standard normalization of the kinetic term of the scalars, we must replace the matrices $M_{e}$, $M_{u}$, $M_{d}$ by
\bbb
Y_{e}&=&
\frac{\pi}{\sqrt{F_{4}(\t M_{e}M_{e}^{*}+3\t M_{d}M_{d}^{*})}}\;M_{e} ,\n\\
Y_{u}&=&
\frac{\pi}{\sqrt{F_{4}(3\t M_{u}M_{u}^{*})}}\; M_{u},\n\\
Y_{d}&=&
\frac{\pi}{\sqrt{F_{4}(\t M_{e}M_{e}^{*}+3\t M_{d}M_{d}^{*})}}\;M_{d}.\n
\eee
To obtain the scalar potential, we have to sum over all loops of the previous diagram. Because of the symmetry between particles and antiparticles, we restrict ourself to the former, bearing in mind that the final result has to be multiplied by two. The resulting quartic potential $V(\Phi_{d},\Phi_{u})$ is,
\bbb
&V(\Phi_{d},\Phi_{u})=
-\frac{1}{2}\mu_{d}^{2}\t(\Phi_{d}\Phi_{d}^{*}) 
-\frac{1}{2}\mu_{u}^{2}\t(\Phi_{u}\Phi_{u}^{*})&\n\\
&+\lambda_{d}\t(\Phi_{d}\Phi_{d}^{*}\Phi_{d}\Phi_{d}^{*})
+\lambda_{u}\t(\Phi_{u}\Phi_{u}^{*}\Phi_{u}\Phi_{u}^{*})
+\lambda_{ud}\t(\Phi_{d}\Phi_{d}^{*}\Phi_{u}\Phi_{u}^{*}),&\n
\eee
with 
\bbb
\mu_{d}=\mu_{u}&=&\sqrt{\frac{2F_{2}}{F_{4}}}\Lambda,\n\\
\lambda_{d}&=&
\frac{\pi^{2}}{F_{4}}
\frac{\t M_{e}M_{e}^{*}M_{e}M_{e}^{*}+3\t M_{d}M_{d}^{*}M_{d}M_{d}^{*}}
{(\t M_{e}M_{e}^{*}+3\t M_{d}M_{d}^{*})^{2}},\n\\
\lambda_{u}&=&
\frac{\pi^{2}}{F_{4}}
\frac{\t M_{u}M_{u}^{*}M_{u}M_{u}^{*}}
{3(\t M_{u}M_{u}^{*})^{2}},\n\\
\lambda_{ud}&=&
\frac{\pi^{2}}{F_{4}}
\frac{\t M_{u}M_{u}^{*}M_{d}M_{d}^{*}}
{\t M_{u}M_{u}^{*}(\t M_{e}M_{e}^{*}+3\,\t M_{d}M_{d}^{*})}.\n
\eee
At this point it seems wothwhile to notice that the last interaction term disappears in the case of the standard model. Indeed, in this case the doublet $\Phi_{u}$ is the quaternionic conjugate of $\Phi_{d}$, so that they are orthogonal and their interaction term vanishes.

\par

Let us now study the spontaneous symmetry breaking. The previous potential depends on the square norms of the fields $|\phi_{u}|^{2}$ and $|\phi_{d}|^{2}$ and also on the modulus of their scalar product $|\t(\Phi_{u}\Phi_{d}^{*})|^{2}$. Since the coefficient $\lambda_{ud}$ is positive, the potential can be minimal only if the scalar product of the two doublets vanishes. By a unitary transformation in $U(2)$, one can always assume that the vaccum expectation values of the fields are given by
$$
V_{u}=\pp{v_{u}\cr 0},\>\>\>\>V_{d}=\pp{0\cr v_{d}}.
$$
The positive numbers $v_{u}$ and $v_{d}$ are determined such that the previous two doublets are actually the minima of the previous potential. This yields the standard formulas
$$
v_{u}=\frac{\mu_{u}^{2}}{4\lambda_{u}},\>\>\>\>
v_{d}=\frac{\mu_{d}^{2}}{4\lambda_{d}}.
$$
These vacuum expectation values clearly break all gauge symmetry expect the electromagnetic one. To find the spectrum of the theory, one expands the scalar fields around the vacuum expectation values
$$
\Phi_{u}=\pp{v_{u}+h_{u}^{0}\cr h_{u}^{\pm}},\>\>\>\>
\Phi_{d}=\pp{h_{d}^{\pm}\cr v_{d}+h_{d}^{0}}.
$$
Expansion of the Higgs potential in terms of the new complex variables $h_{u}^{0}$, $h_{d}^{0}$, $h_{u}^{0}$ and $h_{u}^{0}$ yields the masses of the physical fields. In the neutral sector $h_{u}^{0}, h_{d}^{0}$, we get two Goldstone bosons corresponding to the broken symmetries associated to the gauge bosons $Z$ and $Z'$, and two Higgs bosons with mass $\sqrt{2F_{2}/F_{4}}\Lambda$. The mass term of the charged sector is given by
$$
\frac{1}{2}\frac{F_{2}}{F_{4}}\Lambda^{2}
\pp{h_{u}^{\pm} &  h_{d}^{\pm}}^{*}
\pp{\frac{\lambda_{ud}}{\lambda_{u}}&
\frac{\lambda_{ud}}{\sqrt{\lambda_{u}\lambda_{d}}}\cr
\frac{\lambda_{ud}}{\sqrt{\lambda_{u}\lambda_{d}}}&
\frac{\lambda_{ud}}{\lambda_{d}}}
\pp{h_{u}^{\pm}\cr h_{d}^{\pm}}
$$
This matrix has a zero mode corresponding to the two Goldstone modes associated to the charge gauge bosons. It also has an other eigenvalues whose eigenvector is a complex scalar fields that corresponds to two charged Higgs particles with mass $m_{\pm}$ satisfying
$$
m_{\pm}^{2}=\frac{F_{2}}{F_{4}}\lambda_{ud}
\lp\frac{1}{\lambda_{u}}+\frac{1}{\lambda_{d}}\rp\,\Lambda^{2}.
$$
In the limit where we neglect all fermionic masses except the top and the bottom ones, we get
$$
\lambda_{u}=\lambda_{d}=\lambda_{ud}=\frac{\pi^{2}}{3F_{4}},
$$
and 
$$
m_{\pm}=\sqrt{\frac{2F_{2}}{F_{4}}}\Lambda.
$$
Therefore, in this theory, all the Higgs scalars share the same mass, which is equal to the mass of the Higgs boson of the standard model, computed in the framework of noncommutative geometry. Contrary to the Connes-Lott model, such an extension of the standard model may be acceptable, provided the additional neutral gauge boson is eliminated.

\section{Conclusion}

Finite spectral triples are the simplest examples of noncommutative geometries; since their algebras of coordinates is finite dimensional, they must be thought of as describing the noncommutative analogue of a finite set of points. To each of these points is associated a simple factor of the algebra, and the gauge field constructed out of the spectral triple is interprated as a connection linking these two points. When tensorized with the usual geometry of a riemannian manifold, a finite spectral triple gives rise to a Yang-Mills-Higgs model coupled to gravity, whose lagrangian is derived from the spectral action principle.

\par

This lagrangian contains the standard Einstein-Hilbert term, a cosmological term and some terms pertaining to $\rrr^{2}$-gravity, that reduce, in dimension four, to the square of the Weyl tensor. Moreover, it provides us with a Yang-Mills theory with spontaneous symmetry breaking. The scalar potential is obtained within a diagrammatic aproach, as a sum over all closed loops of the diagram associated to the finite spectral triple. It turns out that all scalar fields share the quadratic term, that is always of order of the cut-off scale $\Lambda$. Besides, there is a general inequality, scale and model independent, between the mass of the heaviest gauge boson and the heaviest fermion. However, all the masses and the coupling depends on $\Lambda$. Following the viewpoint of \cite{spec}, these relations are considered as valid at the Planck scale and results at the electroweak scale should be obtained through the renormalization flow.

\par

The methods we have developped in the previous sections have been illustrated on a simple extension of the standard model, whose scalar sector is more involved. In this case, we get the Higgs potential as a sum over loops, with two complex doublets of self-interacting scalar fields and also a non trivial interaction term between themselves, leading to four physical Higgs bosons with masses of order of the cut-off scale $\Lambda$.

\vskip 0.5truecm
\noindent

{\Large\bf Aknowledgements}\\

Professors B. Iochum, D. Kastler and T. Sch\"ucker are warmly thanked for their unvaluable help and advices.

\end{document}